\documentclass[singlespacing]{elsart}
\usepackage{graphicx}
\usepackage{amssymb} 
\begin{document}

\begin{frontmatter}
\title{Limits on the WIMP-Nucleon Coupling Coefficients from Dark Matter Search Experiment with NaF Bolometer}
\author[1]{\corauthref{cor1}\thanksref{0}A.~Takeda},
\ead{takeda@cr.scphys.kyoto-u.ac.jp}
\thanks[0]{{\it Present Address:} Cosmic-Ray Group, Department of Physics, Faculty of Science, Kyoto University, Kitashirakawa, Sakyo-ku, Kyoto 606-8502, Japan}
\author[1,2]{M.~Minowa},
\author[4]{K.~Miuchi},
\author[1]{H.~Sekiya},
\author[1]{Y.~Shimizu},
\author[2,3]{Y.~Inoue},
\author[3]{W.~Ootani},
\author[5]{Y.~Ootuka}
\address[1]{Department of Physics, School of Science, University of Tokyo, 7-3-1, Hongo, Bunkyo-ku, Tokyo 113-0033, Japan}
\address[2]{Research Center for the Early Universe(RESCEU), School of Science, University of Tokyo, 7-3-1, Hongo, Bunkyo-ku, Tokyo 113-0033, Japan}
\address[3]{International Center for Elementary Particle Physics(ICEPP), University of Tokyo, 7-3-1, Hongo, Bunkyo-ku, Tokyo 113-0033, Japan}
\address[4]{Cosmic-Ray Group, Department of Physics, Faculty of Science, Kyoto University, Kitashirakawa, Sakyo-ku, Kyoto, 606-8502, Japan}
\address[5]{Institute of Physics, University of Tsukuba, 1-1-1 Ten'nodai, 
Tsukuba, Ibaraki 305-8571, Japan}
\corauth[cor1]{Corresponding author.}
\begin{abstract}
We have performed the underground dark matter search experiment with a sodium fluoride (NaF) bolometer array from 2002 through 2003 at Kamioka Observatory (2700 m.w.e.). 
The bolometer array consists of eight NaF absorbers with a total mass of 176 g, and sensitive NTD germanium thermistors glued to each of them.
This experiment aims for the direct detection of weakly interacting massive particles (WIMPs) via spin-dependent interaction.
With an exposure of 3.38 ${\rm kg \, days}$, we derived the limits on the WIMP-nucleon coupling coefficients, $a_{\rm p}$ and $a_{\rm n}$.
These limits confirmed and tightened those derived from our previous results with the lithium fluoride (LiF) bolometer.
Our results excluded the parameter space complementary to the results obtained by NaI detectors of UKDMC experiment.
\end{abstract}
\end{frontmatter}

\newpage
\section{Introduction}
\label{intro}
Several lines of evidence indicate that the universe contains a large amount of nonbaryonic dark matter.
The most plausible candidates for this nonbaryonic dark matter are weakly interacting massive particles (WIMPs) provided by the supersymmetry in the form of the lightest supersymmetric particles (LSPs).
LSPs can be directly detected through their elastic scattering off target nuclei in a detector \cite{MWGoodman}.
Many experiments have been performed for the direct detection of WIMPs.
Among them, the DAMA experiment has reported an annual modulation signal from the four years' measurement with nine 10 kg NaI(Tl) detectors \cite{DAMA_annual}.
The parameters reported by the DAMA experiment are $M_{\rm WIMP}=(52^{+10}_{-8})$ GeV for the WIMP mass and $\sigma^{\rm SI}_{\rm WIMP-p} = (7.2^{+0.4}_{-0.9}) \times 10^{-6}$ pb for the WIMP-proton spin-independently (SI) coupled cross section.
On the other hand, the CDMS \cite{CDMS_2000} and EDELWEISS \cite{EDELWEISS_2002} experiments have excluded most of the parameter space claimed by the DAMA experiment under the assumptions of standard WIMP interactions and a halo model.
Since the cross section for the WIMP-nucleus elastic scattering is given by the sum of SI and spin-dependent (SD) terms, the dark matter search experiments should be performed both via SI and SD interactions.
In an SI and SD mixed framework~\cite{DAMA_SDSI}, $\sigma^{\rm SD}_{\rm WIMP-p}$ could be as high as 0.6 pb for 30--80 GeV WIMPs in the DAMA allowed region under the assumptions of certain parameter sets.
Since the limits for SD WIMP-proton cross section,  $\sigma^{\rm SD}_{\rm WIMP-p}$, obtained by the conventional method \cite{PFSmith} are highly WIMP model dependent, both $\sigma^{\rm SD}_{\rm WIMP-p}$ and SD WIMP-neutron cross section, $\sigma^{\rm SD}_{\rm WIMP-n}$, limits or the limits in terms of the fundamental WIMP-nucleon coupling coefficients $a_{\rm p}$ and $a_{\rm n}$ should be derived \cite{Tovey}.

We, the Tokyo group, reported our previous results with the lithium fluoride (LiF) bolometer at Kamioka Observatory (2700 m.w.e.) \cite{Tokyo_2002}, and set the new limits in the $a_{\rm p}$--$a_{\rm n}$ plane.
We compared our results with those of the UKDMC experiments \cite{UK_1996,UK_2000}, and tightened the UKDMC's limits that are close to the DAMA's SD allowed region in Ref.~\cite{DAMA_SDSI}.

Recently, we developed a sodium fluoride (NaF) bolometer array, and performed a dark matter search measurement from 2002 through 2003.
Since ${\rm ^{23}Na}$ has a larger expectation value of the neutron spins than ${\rm ^{7}Li}$ as shown in Table \ref{TableSpSn}, we were able to tighten the limits from the previous experiment \cite{Tokyo_2002} in the $a_{\rm p}$--$a_{\rm n}$ plane.

In this paper, we present the results from a dark matter search experiment performed at Kamioka Observatory using the NaF bolometer array. 

\section{Experimental set-up}
\label{set-up}
The present data were collected at the Kamioka Observatory under a 2700 m.w.e. rock overburden located at Mozumi Mine of the Kamioka Mining and Smelting Co. in Kamioka-cho, Gifu, Japan.
Details of the underground laboratory and the experimental set-up are given in Ref.~\cite{Tokyo_2002}.

The detector is an array of eight NaF bolometers with a total mass of 176 g.
Each bolometer consists of $2 \times 2 \times 2 \, {\rm cm^{3}}$ NaF crystal and a neutron transmutation doped (NTD) germanium thermistor glued to it.
Each NTD thermistor is biased through a 300 M$\Omega$ load resistor, and the voltage change across the thermistor is fed to the voltage sensitive amplifier with a gain of 10 placed at the 4 K cold stage.
The signal from the cold stage amplifier is in turn amplified by the voltage amplifier with a gain of $4.86 \times 10^{2}$ placed at the room temperature, and is digitized with a sampling rate of 1 kHz.
All the data are recorded continuously without any online trigger for a complete offline analysis.

We refer to each bolometer as from D1 to D8.
Five of the NaF crystals are purchased from OHYO KOKEN KOGYO Co. Ltd., two of them are from RARE METALLIC Co. Ltd., and the other is from MATERIAL TECHNOLOGY Co. Ltd.
We used the NTD thermistors, which are referred to as NTD3 in Ref.~\cite{NTD}, for the eight bolometers.
The concentration of radioactive contaminations in the NaF crystals was previously checked with a low-background germanium spectrometer placed at Kamioka Observatory (2700 m.w.e.), and was found to be less than 0.1 ppb for U, 1 ppb for Th, and 2 ppm for K. 
In order to eliminate remaining grains of the polishing powder, surfaces of the NaF crystals were etched with ultra-pure water in a clean bench of class 100.

The bolometer array is encapsulated in an inner lead shield with a thickness of 2 cm.
In the previous measurement, this inner shield was made of over 200 years old lead.
We used a new inner shield with a low-radioactivity lead older than 400 years for this experiment.

\section{Measurement and analysis}
From 23rd December 2002 through 24th January 2003, a dark matter search measurement was performed.
In this work, D6 and D7 showed poor signal to noise ratios due to the incomplete thermal conduction between the crystal and the thermistor and due to the high electric noises, respectively.
We therefore analyzed the data obtained with the other six bolometers.

We performed numerical pulse shaping to all the data using the optimal filter, or the Wiener filter ~\cite{WFilter}.
The energy of each event was determined by the peak height of a pulse after shaping.
The shaped waveform $U(\omega)$ in the Fourier space is given by
\begin{eqnarray}
U(\omega) = \frac{S^{*}(\omega)C(\omega)}{|N(\omega)|^{2}},
\end{eqnarray}
where $S(\omega)$, $C(\omega)$ and $N(\omega)$ are Fourier transformations of the ideal signal waveform, the measured waveform and the noise, respectively, and the asterisk denotes complex conjugation.
Averaged signal of the gamma-ray source of ${\rm ^{137}Cs}$ was used as the ideal signal waveform, while the noises were obtained by gathering data while no pulse, whose height above a sufficiently low level, exists.
A good lineality of this shaping method was confirmed using simulated pulses which were obtained by adding the ideal signals of known energy to the raw data.
Since the signals shaped from the real events have the symmetrical waveform, the asymmetrical waveform from any electric noise events or any events that were thought to have occurred within the thermistors were eliminated by the waveform analysis.
The event selection efficiency was estimated by analyzing the simulated pulses in the same manner as the real data because the deficiency is mainly due to the threshold smeared by the baseline fluctuation caused by the noises.
Energy resolution of each detector was also estimated by a similar method, and listed in Table \ref{DetectorResponse}.
Energy threshold of each detector was set at the energy in which the efficiency was equal to 0.8, and the events above this threshold were selected as real bolometer signals.
Finally, those events which have associated signals in other bolometers within $\pm10$ ms were discarded because the expected rate of the coincident WIMP events is extremely low.

Calibrations were carried out using the ${\rm ^{60}Co}$ and ${\rm ^{137}Cs}$ gamma-ray sources by the same manner as described in Ref.~\cite{Tokyo_2002}.
Obtained gains for six bolometers are listed in Table \ref{DetectorResponse}.
Gains and linearities were checked every four or five days, and the gains were found to be stable within $\pm5\%$.

\begin{table}[htbp]
\begin{center}
\begin{tabular}{ l cccccc}
\hline
Detector                 & D1   & D2  & D3   &D4    & D5   & D8    \\
\hline
Bias Current [nA]        &0.05  &0.07 &0.21  &0.04  &0.04  &0.04   \\ 
Gain [mV/keV]            &1.4   &1.8  &0.88  &0.80  &1.3   &0.95   \\
Threshold [keV]          &14.5  &17.2 &11.6  &37.1  &17.8  &28.0   \\
Event rate [$10^{-3}$Hz] &1.1   &0.51 &0.65  &1.0   &0.95  &1.3    \\
Resolution (FWHM)[keV]   &1.9   &3.2  &1.4   &6.4   &2.8   &5.4    \\ 
Mass [g]                 &21.5  &21.5 &21.7  &21.6  &21.5  &23.4   \\
Live Time [days]         &26.0  &25.2 &26.0  &26.0  &25.5  &25.9   \\
\hline
\end{tabular}
\caption{Detector responses of the six bolometers.}
\label{DetectorResponse}
\end{center}
\end{table}

\begin{figure}[p]
  \begin{center}
  \includegraphics[width=0.7\linewidth]{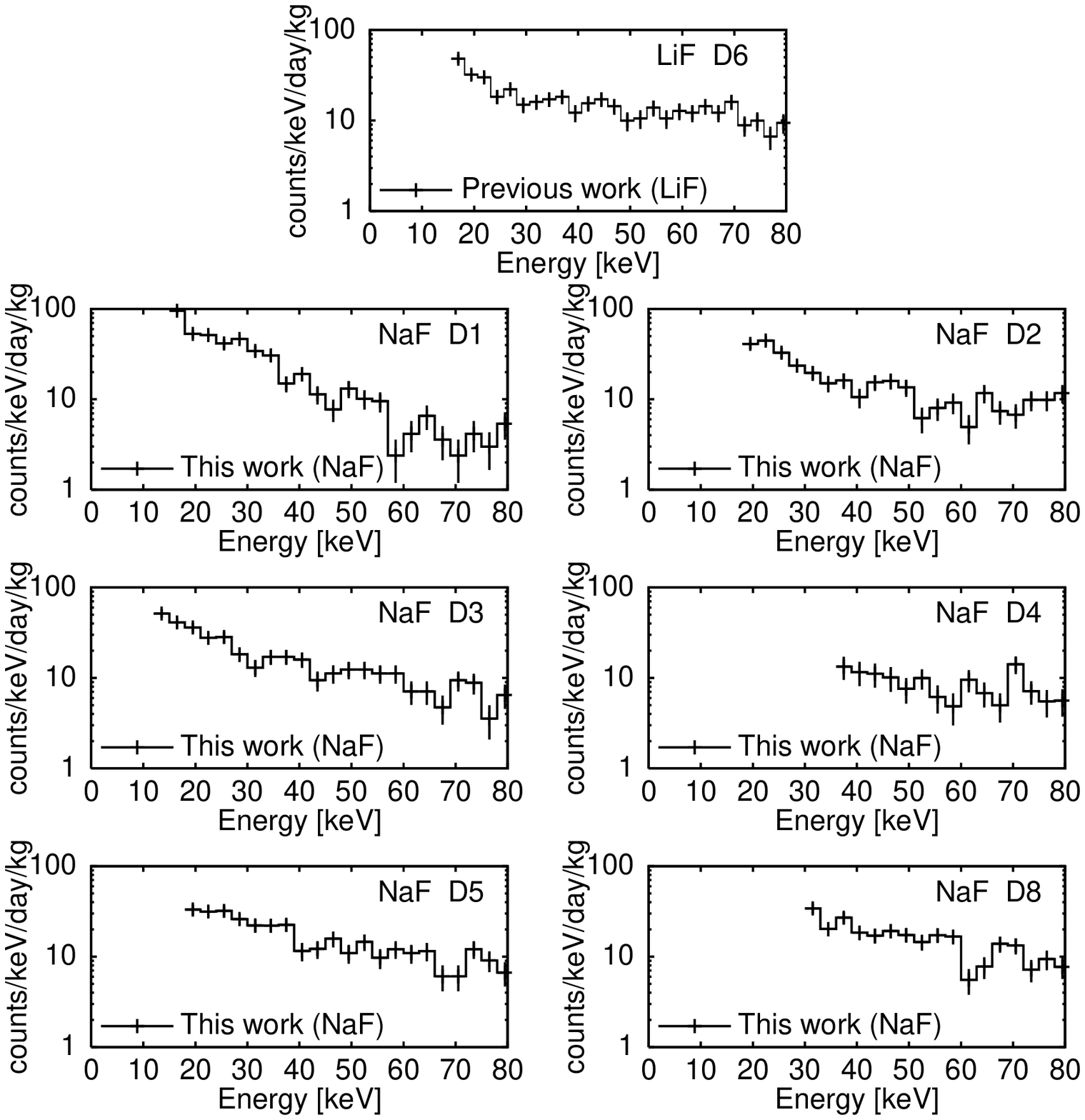}
  \caption{Low energy spectra obtained with the six NaF bolometers are shown. One of the spectra (D6) obtained in the previous measurement with the LiF bolometer \cite{Tokyo_2002} is also shown on top.}
  \label{FigAllSpectraLow}
  \end{center}
\end{figure}

Fig. \ref{FigAllSpectraLow} shows the spectra obtained with the six bolometers together with one of the spectra (D6) obtained in the previous measurement with the LiF bolometer \cite{Tokyo_2002}.
Live times of the six bolometers are listed in Table \ref{DetectorResponse}.
The total exposure is 3.38 ${\rm kg \, days}$.
From this measurement with the NaF bolometer, we obtained the same background levels as the previous measurement.

\begin{figure}[p]
  \begin{center}
  \includegraphics[width=.6\linewidth]{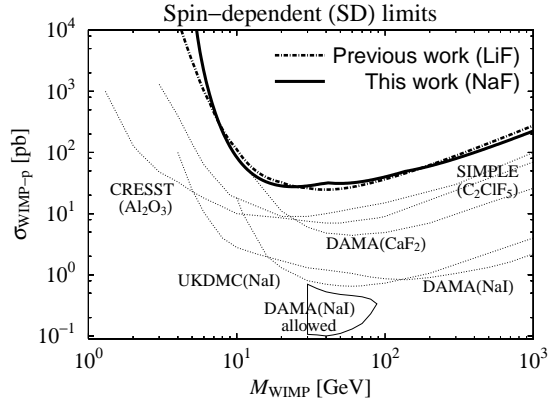}
   \caption{NaF combined 90\% C.L. $\rm \sigma^{\rm SD}_{\rm WIMP-p}$ limits as a function of $M_{\rm WIMP}$ are shown in a thick solid line. Limits from our previous results with the LiF bolometer at Kamioka Observatory \cite{Tokyo_2002} and from other experiments \cite{UK_2000,Tokyo_1999_PLB,DAMA_limit,DAMA_CaF2,SIMPLE,CRESST} are shown in a thick dash-dotted and thin dotted lines, respectively. DAMA's allowed region investigated in a mixed coupling framework shown as ``case c'' in Ref. \cite{DAMA_SDSI} is shown in a closed contour. }
  \label{FigLimitsSD}
  \end{center}
\end{figure}

From the obtained spectra, we derived the limits on the SD WIMP-proton elastic scattering cross section, $\sigma^{\rm SD}_{\rm WIMP-p}$, in the same manner as used in Refs. \cite{PFSmith,Tokyo_2002}.
The limits are shown in Fig. \ref{FigLimitsSD}.
The best limit is 27 pb for $M_{\rm WIMP} = 20 \, {\rm GeV}$.
Here we used the astrophysical and nuclear parameters listed in Table \ref{TableValuesForLimit}.
We used the $\lambda^{2}J(J+1)$ values calculated by assuming the odd group shell model \cite{Ellis93}, where $\lambda$ is $\rm Land\acute{e}$ factor and $J$ is the total spin of the nucleus.

\begin{table}[htbp]
\begin{center}
\begin{tabular}{l l}
\hline
Dark matter density   & $\rho_{\rm D}=\rm 0.3 \, GeVc^{-2}/cm^{3}$ \\ 
Velocity distribution & Maxwellian \\ 
Velocity dispersion   & $v_0$ = 220 $\rm kms^{-1}$ \\ 
Escape velocity       & $v_{\rm esc}$ = 650$\rm kms^{-1}$ \\ 
Earth velocity        & $v_{\rm E}$ = 217 $\rm kms^{-1}$ \\ \hline
Spin factor ($\rm {}^{23}Na$)& $\lambda^2J(J+1)$ = 0.041\\ 
Spin factor ($\rm {}^{19}F$) & $\lambda^2J(J+1)$ = 0.647\\ \hline
\end{tabular}
\caption{Astrophysical and nuclear parameters used to derive the exclusion limit.}
\label{TableValuesForLimit}
\end{center}
\end{table}

As discussed previously, these $\sigma^{\rm SD}_{\rm WIMP-p}$ limits are fraught with potentially significant WIMP model dependence.
We then derived the model-independent limits on $\sigma^{\rm SD}_{\rm WIMP-p}$ and $\sigma^{\rm SD}_{\rm WIMP-n}$ in the same procedure as described in Ref. \cite{Tovey}.
The SD WIMP-nucleus cross section, $\sigma^{\rm SD}_{\rm WIMP-N}$, is given by the following equation:
\begin{eqnarray}
\sigma^{\rm SD}_{\rm WIMP-N} & = & 4 G^{2}_{\rm F} \mu^{2}_{\rm WIMP-N} \times \frac{8}{\pi}(a_{\rm p} \langle S_{\rm p(N)} \rangle + a_{\rm n} \langle S_{\rm n(N)} \rangle)^{2} \frac{J+1}{J},\\
 & = & 4 G^{2}_{\rm F} \mu^{2}_{\rm WIMP-N} \times \left(\sqrt{C^{\rm SD}_{\rm p(N)}} \pm \sqrt{C^{\rm SD}_{\rm n(N)}} \right)^{2} \label{EqSDsigma},
\end{eqnarray}
where $G_{\rm F}$ is the Fermi coupling constant, $\mu_{\rm WIMP-N}$ is the WIMP-nucleus reduced mass, $a_{\rm p}$ and $a_{\rm n}$ are the WIMP-proton and WIMP-neutron coupling coefficients, respectively, $\langle S_{\rm p(N)} \rangle$ and $\langle S_{\rm n(N)} \rangle$ are the expectation values of the proton and neutron spins within the nucleus N, respectively, $C^{\rm SD}_{\rm p(N)}$ and $C^{\rm SD}_{\rm n(N)}$ are the proton and neutron contributions to the total enhancement factor of nucleus N, respectively.
The sign in the square of Eq. (\ref{EqSDsigma}) is determined by the relative sign of $a_{\rm p} \langle S_{\rm p(N)} \rangle$ and $a_{\rm n} \langle S_{\rm n(N)} \rangle$.
$\langle S_{\rm p(N)} \rangle$ and $\langle S_{\rm n(N)} \rangle$ calculated with recent shell models are shown in Table \ref{TableSpSn}.
From the WIMP-nucleus cross section limit, $\sigma^{\rm SD \, lim}_{\rm WIMP-N}$, which is directly set by the experiments, the respective limits on WIMP-proton and WIMP-neutron cross section, $\sigma^{\rm SD \, lim}_{\rm WIMP-p(N)}$ or $\sigma^{\rm SD \, lim}_{\rm WIMP-n(N)}$, are given by assuming that all events are due to WIMP-proton and WIMP-neutron elastic scatterings in the nucleus, such as:
\begin{equation}
\sigma^{\rm SD \, lim}_{\rm WIMP-p(N)} = \sigma^{\rm SD \, lim}_{\rm WIMP-N}\frac{\mu^{2}_{\rm WIMP-p}}{\mu^{2}_{\rm WIMP-N}}\left( \frac{C^{\rm SD}_{\rm p(N)}}{C^{\rm SD}_{\rm p}} \right)^{-1},
\end{equation}
\begin{equation}
\sigma^{\rm SD \, lim}_{\rm WIMP-n(N)} = \sigma^{\rm SD \, lim}_{\rm WIMP-N}\frac{\mu^{2}_{\rm WIMP-n}}{\mu^{2}_{\rm WIMP-N}}\left( \frac{C^{\rm SD}_{\rm n(N)}}{C^{\rm SD}_{\rm n}} \right)^{-1},
\end{equation}
where $\mu_{\rm WIMP-p}$ and $\mu_{\rm WIMP-n}$ are the WIMP-proton and WIMP-neutron reduced mass, respectively, and $C^{\rm SD}_{\rm p}$ and $C^{\rm SD}_{\rm n}$ are the enhancement factors of proton and neutron themselves, respectively.
Ratios of the enhancement factors calculated by the shell models are shown in Table \ref{TableSpSn}.

\begin{table}[b]
  \begin{center}
  \begin{tabular}{ccccccc}
   \hline
  Isotopes & Abundance (\%)   & $J$ & $\langle S_{\rm p(N)}\rangle $&$\langle S_{\rm n(N)}\rangle $&$C^{\rm SD}_{\rm p(N)}/C^{\rm SD}_{\rm p}$&$C^{\rm SD}_{\rm n(N)}/C^{\rm SD}_{\rm n}$ \\
 \hline
    $^7{\rm Li}$ & 92.5 & 3/2 & 0.497& 0.004&$\rm 5.49\times10^{-1}$&$\rm 3.56\times10^{-5}$  \\
    $^{19}{\rm F}$ & 100 &1/2&0.441  & --0.109&$\rm 7.78\times10^{-1}$&$\rm 4.75\times10^{-2}$  \\ 
    $^{23}{\rm Na}$ & 100 &3/2&0.248  & 0.020&$\rm 1.37\times10^{-1}$&$\rm 8.89\times10^{-4}$  \\ 
    $^{127}{\rm I}$ & 100 &5/2&0.309  & 0.075&$\rm 1.78\times10^{-1}$&$\rm 1.05\times10^{-2}$  \\ \hline
        \end{tabular}
\caption{Values of $\langle S_{\rm p(N)}\rangle ,\langle S_{\rm n(N)}\rangle ,C^{\rm SD}_{\rm p(N)}/C^{\rm SD}_{\rm p},C^{\rm SD}_{\rm n(N)}/C^{\rm SD}_{\rm n}$ for various isotopes. Values for ${}^{7}$Li and ${}^{19}$F are taken from Ref.\cite{LiSpSn} and those for ${}^{23}$Na and  ${}^{127}$I are taken from Ref.\cite{127SpSn}}
\label{TableSpSn}
\end{center}
\end{table}

\begin{figure}[p]
  \begin{center}
  \includegraphics[width=0.9\linewidth]{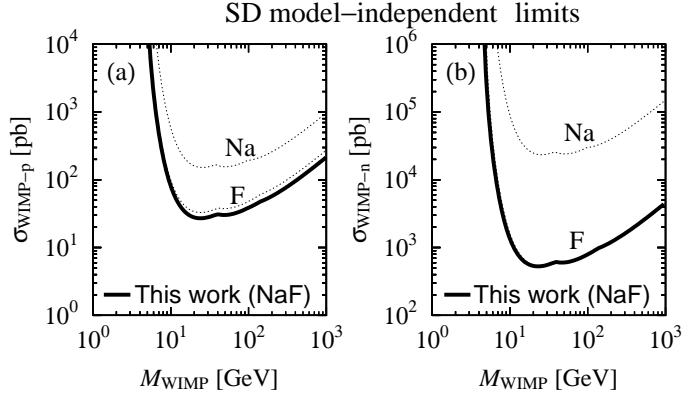}
   \caption{Model independent 90\% C.L. limits on spin-dependent WIMP-proton (a) and WIMP-neutron (b) cross sections as a function of $M_{\rm WIMP}$. The limits have been derived by the same manner as described in Ref. \cite{Tovey}. The limits from Na and F are shown in thin dotted lines. The combined limits are also shown in thick solid lines.}
  \label{FigLimitsMISD}
  \end{center}
\end{figure}

\begin{figure}[p]
  \begin{center}
  \includegraphics[width=0.9\linewidth]{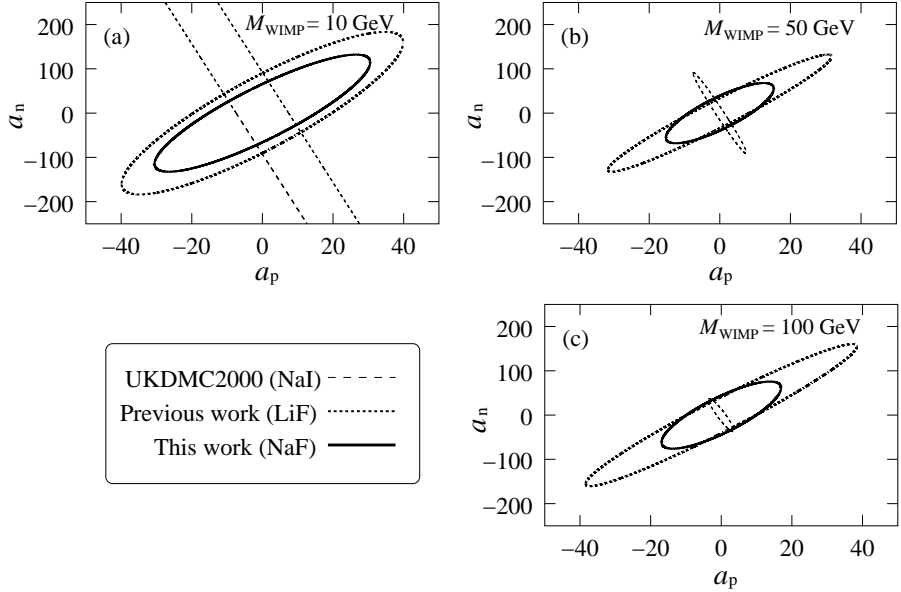}
  \caption{Limits in the $a_{\rm p}$--$a_{\rm n}$ plane: (a) For $M_{\rm WIMP} = 10 \, {\rm GeV}$, (b) for $M_{\rm WIMP} = 50 \, {\rm GeV}$, and (c) for $M_{\rm WIMP} = 100 \, {\rm GeV}$. Inside of thick solid lines are allowed region by this work. The limits from our previous results with the LiF bolometer at Kamioka Observatory and UKDMC data\cite{UK_2000} are also shown for comparison in thick dotted lines and thin dashed lines, respectively.}
  \label{FigLimitapanUK}
  \end{center}
\end{figure}

The obtained model-independent limits on $\sigma^{\rm SD}_{\rm WIMP-p}$ and $\sigma^{\rm SD}_{\rm WIMP-n}$ are shown in Fig. \ref{FigLimitsMISD}.
We then derived the limits in the $a_{\rm p}$--$a_{\rm n}$ plane from them.
According to Ref. \cite{Tovey}, the allowed region in the $a_{\rm p}$--$a_{\rm n}$ plane is defined as:
\begin{equation}
\left( \frac{a_{\rm p}}{\sqrt{\sigma^{\rm SD \, lim}_{\rm WIMP-p(N)}}} \pm \frac{a_{\rm n}}{\sqrt{\sigma^{\rm SD \, lim}_{\rm WIMP-n(N)}}} \right)^{2} < \frac{\pi}{24 G^{2}_{\rm F} \mu^{2}_{\rm WIMP-p}},
\end{equation}
where the relative sign inside the square is determined by the sign of $\langle S_{\rm p(N)} \rangle / \langle S_{\rm n(N)} \rangle$, and the small proton-neutron mass difference is ignored.
Obtained limits in the $a_{\rm p}$--$a_{\rm n}$ plane for WIMPs with mass of 10 GeV, 50 GeV and 100 GeV are shown in Fig. \ref{FigLimitapanUK}.
The limits on the respective coupling coefficients are, for instance, $|a_{\rm p}| < 16$ and $|a_{\rm n}| < 68$ for $M_{\rm WIMP} = 50 \, {\rm GeV}$.

In Fig. \ref{FigLimitapanUK}, we compared our results with those of our previous results \cite{Tokyo_2002} and those of UKDMC experiment \cite{UK_2000}.
Though our $\sigma^{\rm SD}_{\rm WIMP-p}$ limits derived from this measurement are almost the same as those derived from the previous results with the LiF bolometer, we improved the limits in the $a_{\rm p}$--$a_{\rm n}$ plane significantly.
This is because \nuc{23}{Na} has a larger expectation value of the neutron spins than \nuc{7}{Li}.
It is clearly seen that our results obtained with LiF or NaF detectors set limits complementary to the results obtained with NaI detectors of UKDMC experiment because the spin factor of \nuc{19}{F} are large and the sign of $\langle S_{\rm p(N)} \rangle / \langle S_{\rm n(N)} \rangle$ for \nuc{19}{F} is opposite to those for \nuc{7}{Li}, \nuc{23}{Na}, and \nuc{127}{I}. 
As described in Ref. \cite{Tokyo_2002}, we did not compare our results with the SD annual modulation allowed regions shown in Fig. \ref{FigLimitsSD} \cite{DAMA_SDSI} because the DAMA group had not shown the results from \nuc{23}{Na} and \nuc{127}{I} independently.

The remaining background in the energy region between 10 and 90 keV restricting the current sensitivity is supposed to be due to the uranium and thorium contamination in a crystal holder made of the OFHC copper.
A use of the crystal holder with a sufficiently radio-pure material such as polytetrafluoroethylene (PTFE) and active shields between the crystals and their holder would help to reduce the background from the holder and the other outside materials.

\section{Conclusion}
\label{discussions_conclusion}
We have performed the dark matter search experiment at Kamioka observatory with the NaF bolometer.
With a total exposure of 3.38 ${\rm kg \, days}$, we have confirmed and tightened the limits in the $a_{\rm p}$--$a_{\rm n}$ plane obtained by the previous results with the LiF bolometer.
For the WIMPs with a mass of 50 GeV, we have successfully excluded more than half of the parameter region in the $a_{\rm p}$--$a_{\rm n}$ plane allowed by the previous results with the LiF bolometer. 
Our results set limits complementary to the results obtained by NaI detectors of UKDMC experiment.
Since the limits from the UKDMC experiments are close to the DAMA's SD allowed region in Ref.~\cite{DAMA_SDSI}, we should also be able to compare our limits in the $a_{\rm p}$--$a_{\rm n}$ plane if DAMA publishes relevant data to calculate these parameters.
It is worth noting that this result has been obtained by a single detector with both ${\rm ^{23}Na}$, which has relatively large $C^{\rm SD}_{\rm n(N)}/C^{\rm SD}_{\rm n}$, and ${\rm ^{19}F}$, which has the largest $C^{\rm SD}_{\rm p(N)}/C^{\rm SD}_{\rm p}$, in the detector material.

\section*{Acknowledgements}
We would like to thank all the staff of Kamioka Observatory of the
Institute for Cosmic Ray Research, the University of Tokyo, to whose
hospitality we owe a great deal in using the facilities of the
Observatory. This research is supported by the Grant-in-Aid for COE Research by the Japanese Ministry of Education, Culture, Sports, Science and Technology.


\newpage
\pagestyle{empty}

\newpage
\pagestyle{empty}

\newpage
\pagestyle{empty}

\newpage
\pagestyle{empty}

\end{document}